\documentclass[twocolumn,aps,showpacs,preprintnumbers,amsmath,amssymb, superscriptaddress]{revtex4-1}
\pdfoutput=1
\usepackage{bm}
\usepackage{graphicx}
\usepackage{color}
\usepackage{xspace}
\definecolor{dblue}{rgb}{0,0,0.6}
\definecolor{dred}{rgb}{0.9,0,0}
\definecolor{dgreen}{rgb}{0,0.4,0}

\begin{document}

\title{Raman scattering investigation of quasi-one-dimensional superconductor Ta$_{4}$Pd$_{3}$Te$_{16}$}

\author{D. Chen}
\affiliation{Beijing National Laboratory for Condensed Matter Physics, and Institute of Physics, Chinese Academy of Sciences, Beijing 100190, China}
\author{P. Richard}\email{p.richard@iphy.ac.cn}
\affiliation{Beijing National Laboratory for Condensed Matter Physics, and Institute of Physics, Chinese Academy of Sciences, Beijing 100190, China}
\affiliation{Collaborative Innovation Center of Quantum Matter, Beijing, China}
\author{Z.-D. Song}
\affiliation{Beijing National Laboratory for Condensed Matter Physics, and Institute of Physics, Chinese Academy of Sciences, Beijing 100190, China}
\author{W.-L. Zhang}
\affiliation{Beijing National Laboratory for Condensed Matter Physics, and Institute of Physics, Chinese Academy of Sciences, Beijing 100190, China}
\author{S.-F. Wu}
\affiliation{Beijing National Laboratory for Condensed Matter Physics, and Institute of Physics, Chinese Academy of Sciences, Beijing 100190, China}
\author{W. H. Jiao}
\affiliation{Department of Physics, Zhejiang University, Hangzhou 310027, China}
\author{Z. Fang}
\affiliation{Beijing National Laboratory for Condensed Matter Physics, and Institute of Physics, Chinese Academy of Sciences, Beijing 100190, China}
\affiliation{Collaborative Innovation Center of Quantum Matter, Beijing, China}
\author{G.-H. Cao}
\affiliation{Department of Physics, Zhejiang University, Hangzhou 310027, China}
\author{H. Ding}\email{dingh@iphy.ac.cn}
\affiliation{Beijing National Laboratory for Condensed Matter Physics, and Institute of Physics, Chinese Academy of Sciences, Beijing 100190, China}
\affiliation{Collaborative Innovation Center of Quantum Matter, Beijing, China}

\date{\today}

\begin{abstract}
We have performed polarized Raman scattering measurements on the newly discovered superconductor Ta$_{4}$Pd$_{3}$Te$_{16}$ ($T_c = 4.6$  K). We observe twenty-eight out of thirty-three Raman active modes, with frequencies in good accordance with first-principles calculations. Although most of the phonons observed vary only slightly with temperature and do not exhibit any asymmetric profile that would suggest strong electron-phonon coupling, the linewidth of the A$_{g}$ phonon mode at 89.9~cm$^{-1}$ shows an unconventional increase with temperature decreasing, which is possibly due to a charge-density-wave transition or the emergence of charge-density-wave fluctuations below a temperature estimated to fall in the 140-200~K range.
\end{abstract}



\maketitle

Because they have anisotropic interactions and electronic properties, low-dimensional systems often show intriguing physical behaviors. Despite a low critical temperature ($T_c$) of 4.6 K \cite{caoguanghan}, the newly discovered quasi-one-dimensional (quasi-1D) superconductor Ta$_{4}$Pd$_{3}$Te$_{16}$ has been labeled as an unconventional superconductor from scanning tunneling microscopy (STM) \cite{STM} and thermal conductivity \cite{thermal} experiments, from which an anisotropic gap structure with possible nodes has been proposed. This result is quite intriguing since the electronic states of typical unconventional superconductors at the Fermi level are derived from $d$ or $f$ bands while a recent density functional theory (DFT) study on Ta$_{4}$Pd$_{3}$Te$_{16}$ indicates that they mainly originate from Te $p$ states in this material with only little contribution from $d$ electrons from transition metal elements \cite{calculation}. In addition, quasi-1D systems are particularly prone to the formation of a static charge-density-wave (CDW) or a static spin-density-wave (SDW), often regarded as states in competition with superconductivity. Not only the electronic structure contains nested Te $p$ bands, which would be favorable to a CDW rather than a SDW \cite{calculation}, a CDW-like gap feature has also been reported from low-temperature STM measurements on stripes of Te atoms \cite{STS}. CDW fluctuations is thus a potential candidate for the pairing glue in Ta$_{4}$Pd$_{3}$Te$_{16}$ that would in principle favor a singlet state \cite{calculation} and strong electron-phonon interactions.  

In order to investigate the strength of the electron-phonon coupling and the possible existence of a CDW instability, here we present a Raman scattering study of Ta$_{4}$Pd$_{3}$Te$_{16}$ supported by first-principles calculations of the vibration modes. Raman scattering is an experimental technique very sensitive to modulations in the lattice that is also capable of probing electron-phonon interactions. We observed twenty-eight out of thirty-three Raman active modes, with frequencies in good accordance with our first-principles calculations. Most of the phonons that we observed suggest no strong electron-phonon coupling and vary only slightly with temperature, except for one Ag phonon at 89.9~cm$^{-1}$. We show that this particular mode, associated to vibrations of Te located along the stripes where a CDW-like gap was observed from STM \cite{STS}, splits in the 140-200 K temperature range and below, which is possibly related to a CDW transition or CDW fluctuations.

\begin{figure}[!t]
\begin{center}
\includegraphics[width=3.4in]{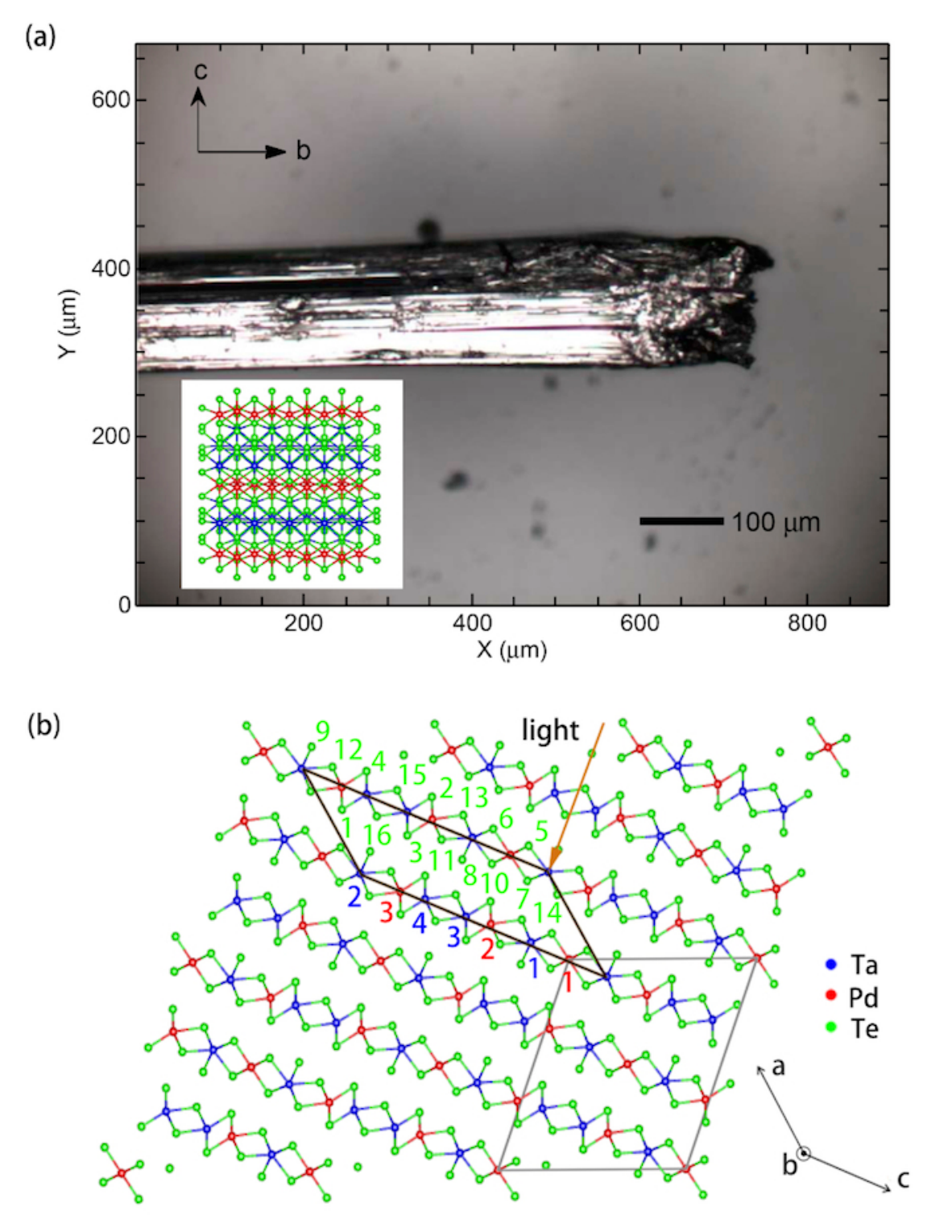}
\end{center}
\caption{\label{fig-1}(Color online). (a) Photo of a Ta$_{4}$Pd$_{3}$Te$_{16}$ single crystal. The inset shows the crystal structure viewed perpendicularly to the sample surface. (b) Crystal structure of Ta$_{4}$Pd$_{3}$Te$_{16}$ viewed perpendicularly to the $b$-axis. The black parallelogram represents the primitive cell used in our Raman measurements, while the grey one refers to another unit cell commonly used. All the atoms in the primitive cell are identified by numbers. The incident light is perpendicular to the $bc$ plane.}
\end{figure}

\begin{table*}
\caption{\label{EXP_CAL_comparsion}Comparison of the calculated and experimental Raman active phonon modes at 294 K.}
\begin{ruledtabular}
\begin{tabular}{ccccc}
 Sym. &	  Exp. &	 Cal.	& Main atom displacements\\
\hline
A$_{g}$&  26.4 &     21.0&  Pd$^2(-xz)$, Pd$^3(xz)$, Te$^9(xz)$, Te$^{10}(-xz)$, Te$^{15}(-xz)$, Te$^{16}(xz)$\\
B$_{g}$&  28.4&	     27.9&	Ta$^1(y)$, Ta$^2(-y)$, Pd$^2(y)$, Pd$^3(-y)$, Te$^8(y)$, Te$^9(-y)$, Te$^{15}(y)$, Te$^{16}(-y)$\\
A$_{g}$&  41.4&      36.9&	Ta$^1(xz)$, Ta$^2(-xz)$, Te$^1(-xz)$, Te$^2(xz)$, Te$^5(xz)$, Te$^6(-xz)$, Te$^{15}(xz)$, Te$^{16}(-xz)$\\
A$_{g}$&  46.8&	     41.6&	Ta$^1(-z)$, Ta$^2(z)$, Pd$^2(xz)$, Pd$^3(-xz)$, Te$^3(xz)$, Te$^4(-xz)$, Te$^{13}(-z)$, Te$^{14}(z)$\\
B$_{g}$&  52.7&      48.2& 	Ta$^1(-y)$, Ta$^2(y)$, Pd$^2(y)$, Pd$^3(-y)$, Te$^5(-y)$, Te$^6(y)$, Te$^7(-y)$, Te$^8(y)$, Te$^9(-y)$, Te$^{10}(y)$\\
A$_{g}$&  58.3&      58.3&	Ta$^3(xz)$, Ta$^4(-xz)$, Pd$^2(-xz)$, Pd$^3(xz)$, Te$^3(z)$, Te$^4(-z)$, Te$^9(-xz)$, Te$^{10}(xz)$, Te$^{15}(-xz)$, Te$^{16}(xz)$\\
B$_{g}$&  61.6& 	 61.5& 	Ta$^3(-y)$, Ta$^4(y)$, Pd$^2(y)$, Pd$^3(-y)$, Te$^3(-y)$, Te$^4(y)$, Te$^{11}(-y)$, Te$^{12}(y)$, Te$^{15}(y)$, Te$^{16}(-y)$\\
A$_{g}$&  65.7& 	 62.1& 	Pd$^2(xz)$, Pd$^3(-xz)$, Te$^3(xz)$, Te$^4(-xz)$, Te$^7(-xz)$, Te$^8(xz)$, Te$^{15}(-z)$, Te$^{16}(z)$\\
A$_{g}$&  74.5&      74.8&	Ta$^3(-xz)$, Ta$^4(xz)$, Te$^1(-xz)$, Te$^2(xz)$, Te$^3(-xz)$, Te$^4(xz)$, Te$^9(xz)$, Te$^{10}(-xz)$\\
A$_{g}$&  81.9&      81.9&	Ta$^3(-xz)$, Ta$^4(xz)$, Te$^5(-z)$, Te$^6(z)$, Te$^{11}(xz)$, Te$^{12}(-xz)$, Te$^9(xz)$, Te$^{10}(-xz)$\\
B$_{g}$&  89.1&      87.2&	Te$^9(-y)$, Te$^{10}(y)$, Te$^{15}(-y)$, Te$^{16}(y)$\\
A$_{g}$&  89.9&      87.8&  Te$^1(-xz)$, Te$^2(xz)$, Te$^5(-xz)$, Te$^6(xz)$, Te$^7(xz)$, Te$^8(-xz)$\\
B$_{g}$&  97.3&      96.5&	Te$^5(y)$, Te$^6(-y)$, Te$^7(-y)$, Te$^8(y)$, Te$^{13}(y)$, Te$^{14}(-y)$\\
A$_{g}$&  99.4&      99.8& 	Te$^1(xz)$, Te$^2(-xz)$, Te$^3(-xz)$, Te$^4(xz)$, Te$^5(z)$, Te$^6(-z)$, Te$^7(xz)$, Te$^8(-xz)$, Te$^{13}(xz)$, Te$^{14}(-xz)$\\
B$_{g}$&  -&         100.6&	Te$^3(y)$, Te$^4(-y)$, Te$^{11}(-y)$, Te$^{12}(y)$\\
B$_{g}$&  103.3&	 101.4&	Te$^1(y)$, Te$^2(-y)$, Te$^5(y)$, Te$^6(-y)$\\
B$_{g}$&  105.2&     105.1&	Te$^1(-y)$, Te$^2(y)$, Te$^7(-y)$, Te$^8(y)$, Te$^{13}(-y)$, Te$^{14}(y)$\\
A$_{g}$&  109.6&     109.4&	Te$^3(xz)$, Te$^4(-xz)$, Te$^5(-xz)$, Te$^6(xz)$, Te$^{13}(-xz)$, Te$^{14}(xz)$\\
A$_{g}$&  -&         116.2& Te$^7(-xz)$, Te$^8(xz)$, Te$^9(-xz)$, Te$^{10}(xz)$, Te$^{15}(xz)$, Te$^{16}(-xz)$\\
A$_{g}$&  120.8&     120.2& Te$^5(xz)$, Te$^6(-xz)$, Te$^7(-z)$, Te$^8(z)$, Te$^{15}(-xz)$, Te$^{16}(xz)$\\
A$_{g}$&  -&         125.2& Pd$^2(xz)$, Pd$^3(-xz)$, Te$^3(-xz)$, Te$^4(xz)$, Te$^{11}(xz)$, Te$^{12}(-xz)$\\
A$_{g}$&  128.8&     125.4& Te$^5(-xz)$, Te$^6(xz)$, Te$^{13}(xz)$, Te$^{14}(-xz)$\\
B$_{g}$&  127.9&     128.5& Ta$^3(y)$, Ta$^4(-y)$, Te$^3(-y)$, Te$^4(y)$\\
A$_{g}$&  139.5&     138.0& Pd$^2(-xz)$, Pd$^3(xz)$, Te$^9(-xz)$, Te$^{10}(xz)$, Te$^{11}(xz)$, Te$^{12}(-xz)$, Te$^{15}(-z)$, Te$^{16}(z)$\\
A$_{g}$&  150.2&     148.7& Pd$^2(xz)$, Pd$^3(-xz)$, Te$^3(-xz)$, Te$^4(xz)$, Te$^{11}(z)$, Te$^{12}(-z)$\\
A$_{g}$&  154.2&     149.6& Te$^1(-xz)$, Te$^2(xz)$, Te$^5(-xz)$, Te$^6(xz)$, Te$^7(-z)$, Te$^8(z)$,\\
B$_{g}$&  -&     157.3& Pd$^2(-y)$, Pd$^3(y)$, Te$^9(-y)$, Te$^{10}(y)$, Ta$^{15}(y)$, Ta$^{16}(-y)$\\
B$_{g}$&  161.2&     161.2& Ta$^1(y)$, Ta$^2(-y)$, Te$^5(-y)$, Te$^6(y)$, Te$^7(-y)$, Te$^8(y)$,\\
A$_{g}$&  168.1&     167.3& Pd$^2(xz)$, Pd$^3(-xz)$, Te$^{15}(-z)$, Te$^{16}(z)$\\
A$_{g}$&  -&         180.6& Ta$^3(-xz)$, Pd$^4(xz)$, Te$^3(-xz)$, Te$^4(xz)$, Te$^9(-xz)$, Te$^{10}(xz)$\\
A$_{g}$&  185.3&     187.3& Ta$^1(z)$, Ta$^2(-z)$, Te$^7(-z)$, Te$^8(z)$\\
A$_{g}$&  192&       193.0& Ta$^1(-xz)$, Ta$^2(xz)$, Te$^5(xz)$, Te$^6(-xz)$, Te$^{15}(xz)$, Te$^{16}(-xz)$\\
A$_{g}$&  206.5&     206.5& Ta$^3(xz)$, Ta$^4(-xz)$, Pd$^2(xz)$, Pd$^3(-xz)$, Te$^3(-xz)$, Te$^4(xz)$, Te$^9(xz)$, Te$^{10}(-xz)$\\

\end{tabular}
\end{ruledtabular}
\begin{raggedright}
The atoms are identified in Figs. \ref{fig-1}(b)\\
\end{raggedright}
\end{table*}

The Ta$_{4}$Pd$_{3}$Te$_{16}$ single crystals used in our Raman scattering measurements were prepared by a self-flux method \cite{caoguanghan}. The shiny, soft, layered samples have a typical dimension of $2.5\times0.1\times0.05$~mm$^{3}$, as shown in Fig. \ref{fig-1}(a). Raman scattering measurements were performed using 488.0~nm, 514.5~nm, 530.9~nm, 568.2~nm and 676.5~nm laser excitations in a back-scattering micro-Raman configuration with a triple-grating spectrometer (Horiba Jobin Yvon T64000) equipped with a nitrogen-cooled CCD camera. At room temperature a $100\times$ objective was used to both focus the laser beam and collect the scattered light whereas a long-focus distance $50\times$ objective was used for measurements in the low-temperature cryostat. The crystals were cleaved by tape to obtain clean and flat surfaces, and then transferred into a low-temperature ST500 (Janis) cryostat for the Raman measurements between 5 and 350 K with a working vacuum better than $2\times 10^{-6}$ mbar. In this paper we use the primitive cell represented by the black line in Fig. \ref{fig-1}(b), which differs from the unit cell used in previous reports \cite{1991synthesis,earliercalculation} on this material (gray line). We define $Y$ and $Z$ as the directions along the $b$ and $c$ axes, respectively. The $Y$ direction corresponds to the PdTe$_{2}$ chains direction. The natural cleaved surface of this crystal is the Ta-Pa-Te layer, \emph{i.e.} the $bc$ plane, as evidenced by STM measurements \cite{STM}.


\begin{figure}[!t]
\begin{center}
\includegraphics[width=3.4in]{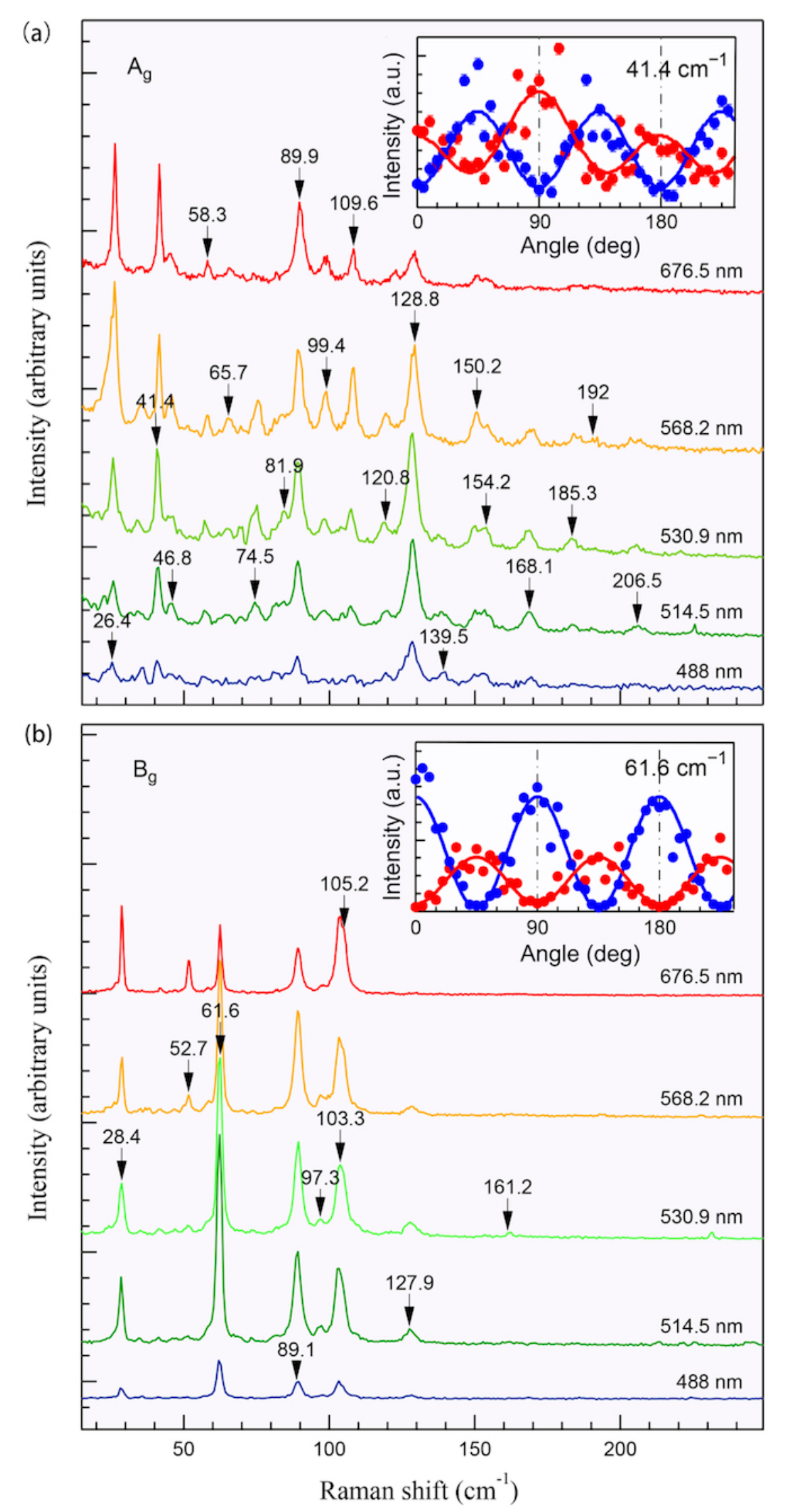}
\end{center}
\caption{\label{fig-2}(Color online). (a) A$_{g}$ modes recorded with 488.0~nm, 514.5~nm, 530.9~nm, 568.2~nm and 676.5~nm laser excitations under the $(YY)$ polarization configuration at room temperature. The curves are shifted relative to each other for clarity. The inset shows the intensity of the peak at 41.4~cm$^{-1}$ as a function of the in-plane angle with respect to the $b$ axis. The red dots and blue dots are measured under parallel and perpendicular polarization configuration, respectively. The blue and red lines are fits of the experimental data based on the A$_g$ tensor. (b) Same as (a) but for the B$_{g}$ modes recorded under the $(YZ)$ polarization configuration at room temperature.}
\end{figure}

The Ta$_{4}$Pd$_{3}$Te$_{16}$ crystal structure is characterized by space group I2/m (point group $C_{2h}$). A simple group symmetry analysis \cite{bilbal} indicates that the phonon modes at the Brillouin zone (BZ) center decompose into [22A$_{g}$+11B$_{g}$]+[11A$_{u}$+22B$_{u}$]+[A$_{u}$+2B$_{u}$], where the first, second and third terms represent the Raman-active modes, the infrared(IR)-active modes and the acoustic modes, respectively. To get estimates on the mode frequencies of Ta$_{4}$Pd$_{3}$Te$_{16}$, we employed the first-principles pseudopotential plane wave method package Quantum Espresso \cite{theory1}. We set a $5\times5\times5$ monkhorst-pack $k$ point mesh and a 40 $Ry$ cutoff for wavefunctions. Using the generalized gradiant approximation \cite{theory2}, we first relax both the cell and atom coordinates of experimental results \cite{1991synthesis} until the forces acting on atoms are all smaller than $10^{-4} Ry/a_B$ and the pressure is smaller than 0.2~kbar. Using the information on the ground state, it is easy to use Phonon package which implements the density functional pertubation theory (DFPT) \cite{theory4,theory5,theory6} to get the A$_g$ and B$_g$ phonon frequencies and vibration modes at the $\Gamma$ point. The calculated phonon modes, their symmetries and the main atomic displacements involved, are given in Table \ref{EXP_CAL_comparsion}.

\begin{figure}[!t]
\begin{center}
\includegraphics[width=3.4in]{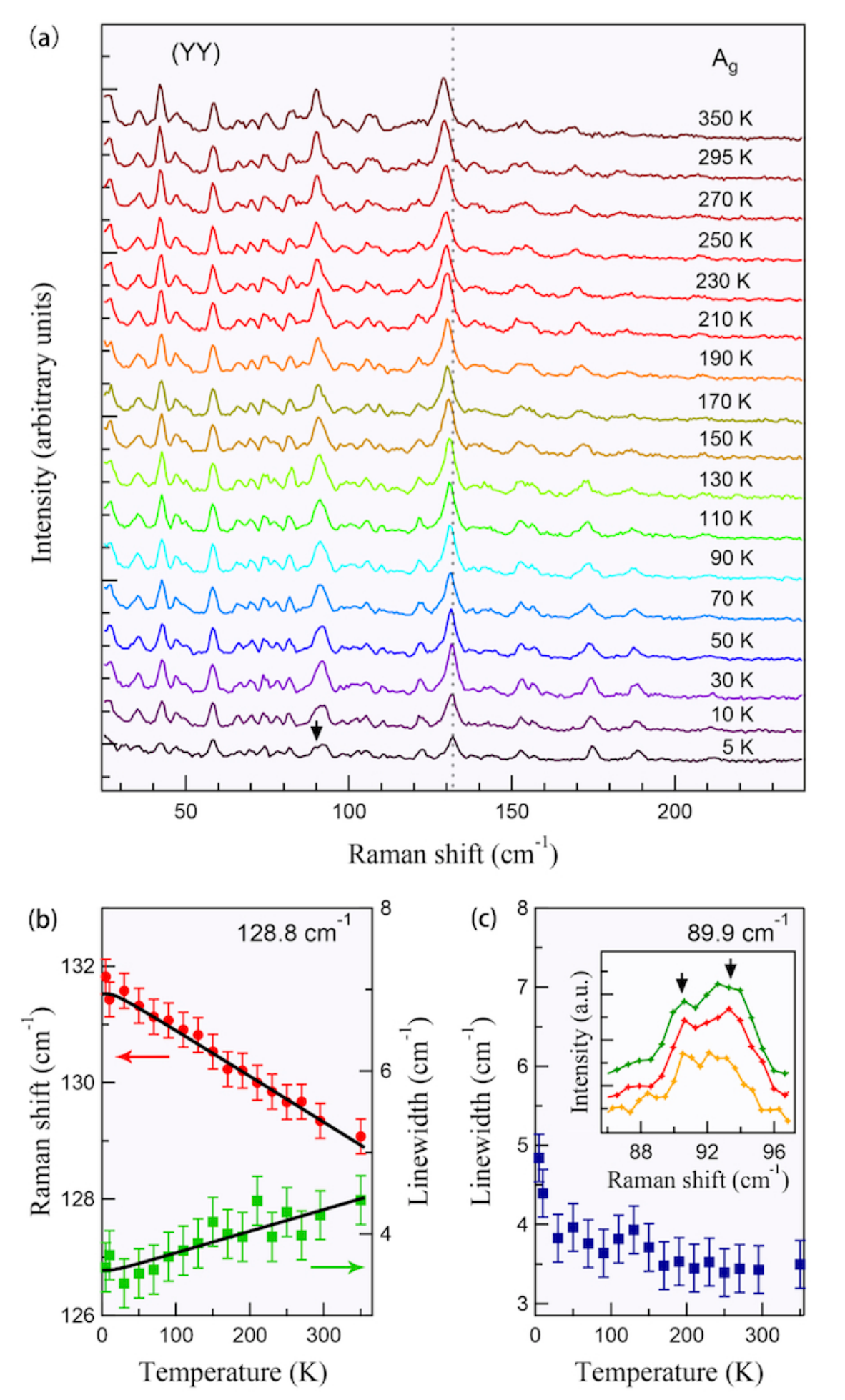}
\end{center}
\caption{\label{fig_Ag}(Color online). (a) Temperature dependence of the Raman spectra for the A$_{g}$ modes. The curves
are shifted relative to each other for clarity. The vertical dashed line and the arrow are guides to eye. (b) Peak position (in red) and linewidth (full width at half maximum, in green), respectively, of the A$_{g}$ mode at 128.8~cm$^{-1}$. The curves are fits to Eqs. \eqref{eq_omega} and \eqref{eq_gamma}. (c) Peak linewidth of the A$_{g}$ mode at 89.9~cm$^{-1}$. The inset shows the spectra of this peak at 5 K. The green spectrum is recorded with 514.5~nm laser excitation. The red and yellow spectra show this peak as measured from another sample with 514.5~nm and 568.2~nm laser excitations, respectively. The arrows suggest the presence of two peak components.}
\end{figure}
\begin{figure}[!t]
\begin{center}
\includegraphics[width=3.4in]{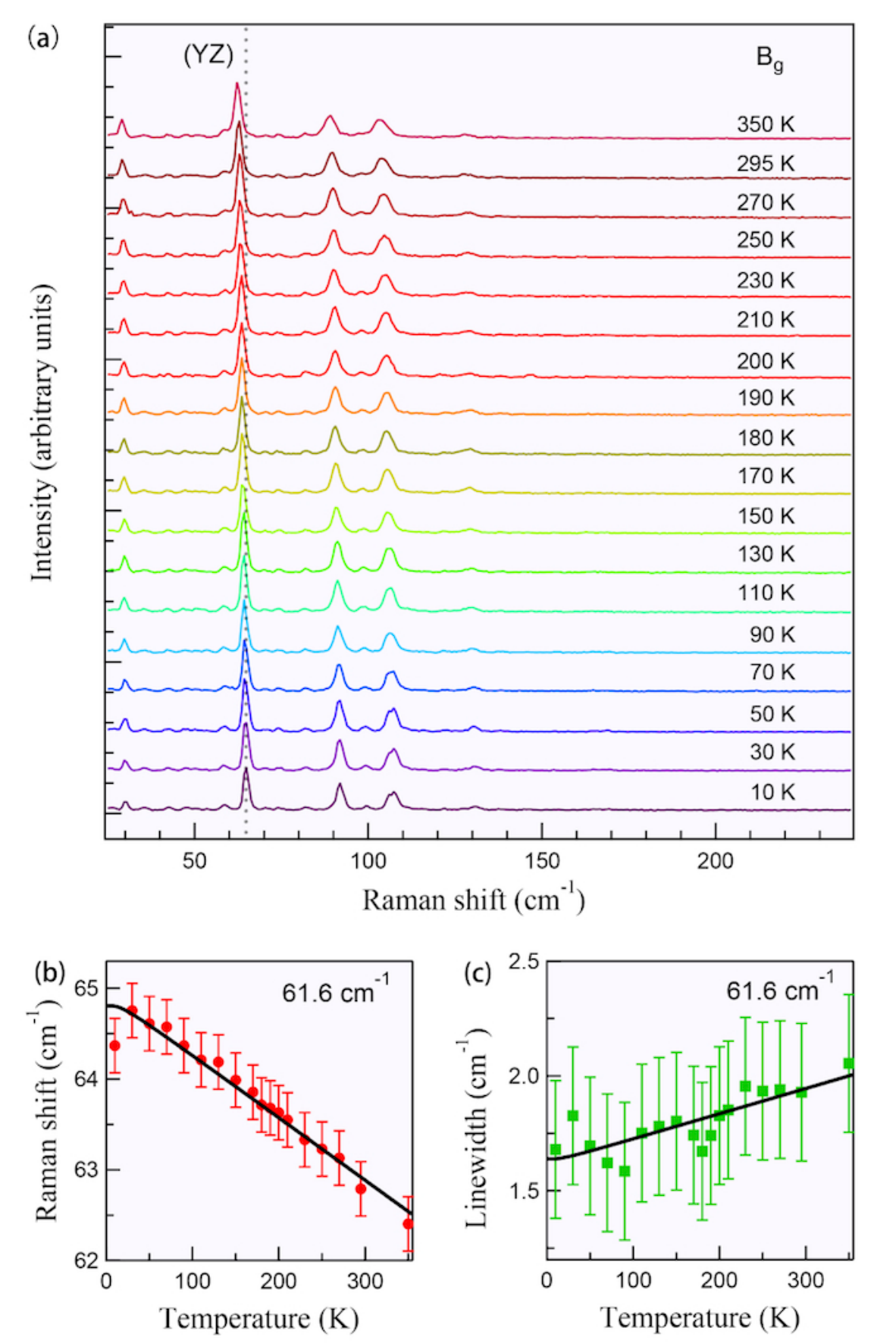}
\end{center}
\caption{\label{fig_Bg}(Color online). (a) Temperature dependence of the Raman spectra for the B$_{g}$ modes. The curves
are shifted relative to each other for clarity. The vertical dashed line is guide to eye. (b)-(c) Peak position and linewidth (full width at half maximum), respectively, of the B$_{g}$ mode at 61.6~cm$^{-1}$. The curves in (b)-(c) are fit to the equations Eqs. \eqref{eq_omega} and \eqref{eq_gamma}.}
\end{figure}

In Figs. \ref{fig-2}(a) and \ref{fig-2}(b), we show the Raman spectra of Ta$_{4}$Pd$_{3}$Te$_{16}$ recorded at room temperature under 488.0~nm, 514.5~nm, 530.9~nm, 568.2~nm and 676.5~nm laser excitations. In total, we observe no less than 28 of the 33 Raman active modes predicted. As indicated in Table \ref{EXP_CAL_comparsion}, the experimental and calculated mode energies are in a very good agreement. Using the polarization selection rules and the fact that the $b$ axis is easily determined by the morphology of the samples, the assignments of the Raman symmetries for each mode is straightforward. The Raman tensors corresponding to the $C_{2h}$ symmetry group are expressed as:

\begin{displaymath}
\textrm{A$_{g}$=}
\left(\begin{array}{ccc}
a & d &0\\
d & b &0\\
0 & 0 &c
\end{array}\right)
, \textrm{B$_{g}$=}
\left(\begin{array}{ccc}
0 & 0 &e\\
0 & 0 &f\\
e & f &0\\
\end{array}\right).
\end{displaymath}

\noindent For perfectly aligned crystals, pure A$_{g}$ symmetry is obtained in the $(YY)$ configuration. In this channel, we detect 19 peaks at 26.4~cm$^{-1}$, 41.4~cm$^{-1}$, 46.8~cm$^{-1}$, 58.3~cm$^{-1}$, 65.7~cm$^{-1}$, 74.5~cm$^{-1}$, 81.9~cm$^{-1}$, 89.9~cm$^{-1}$, 99.4~cm$^{-1}$, 109.6~cm$^{-1}$, 120.8~cm$^{-1}$, 128.8~cm$^{-1}$, 139.5~cm$^{-1}$, 150.2~cm$^{-1}$, 154.2~cm$^{-1}$, 168.1~cm$^{-1}$, 185.3~cm$^{-1}$, 192~cm$^{-1}$ and 206.5~cm$^{-1}$. They correspond to vibrations of atoms in the $bc$ plane. Pure B$_{g}$ symmetry is obtained in the $(YZ)$ configuration. In this channel, we detect 9 peaks at 28.4~cm$^{-1}$, 52.7~cm$^{-1}$, 61.6~cm$^{-1}$, 89.1~cm$^{-1}$, 97.3~cm$^{-1}$, 103.3~cm$^{-1}$, 105.2~cm$^{-1}$, 127.9~cm$^{-1}$, and 161.2~cm$^{-1}$. They correspond to vibrations of atoms along the $b$-axis. The symmetric peak profiles that we observe suggest that the electron-phonon coupling is not particularly strong in this system. 

We also checked how the peak intensities vary away from perfect orientation. Assuming, in a first approximation, that the matrix elements of the Raman tensors are real numbers, the intensity of the A$_{g}$ modes should vary with the angle $\theta$ between the $b$ axis and the incident polarization vector $\mathbf e_i$ as $|b\cos^2 \theta+ c\sin^2 \theta|^2$ and $|(b-c)\sin 2\theta|^2$ for parallel and cross polarization configurations, respectively. For the cross polarization configuration, this implies a four-fold symmetry with maxima at 45 degrees and each successive 90 degrees, which is exactly what we observe for the A$_g$ peak at 41.4~cm$^{-1}$, as shown in the inset of Fig. \ref{fig-2}(a). For the parallel configuration, the expected angular dependence of the A$_g$ peaks depends on the relative sign and intensity of the $b$ and $c$ elements in the Raman tensor. For the peak at 41.4~cm$^{-1}$, we observe small maxima at $\theta=0$ and $\theta=180$ degrees, a larger maxima at $\theta=90$ degrees and four-fold minima starting at $\theta=45$ degrees. This situation is only possible if $b$ and $c$ have a different sign and $|c/b|>$1. Indeed, we found $c/b=-1.5$ from the fitting of the experimental data. The intensity of the B$_{g}$ mode peaks should be proportional to $|f\sin 2\theta|^2$ and $|f\cos 2\theta|^2$ for parallel and cross polarization configurations, respectively. This coincides in both cases with four-fold oscillations, with the first maximum at $\theta=45$ degrees and $\theta=0$ degrees, respectively, in agreement with the results illustrated in the inset of Fig. \ref{fig-2}(b) for the B$_g$ peak at 61.6~cm$^{-1}$. 

To investigate the possible role of the electron-phonon coupling and the possible existence of a CDW order, we cooled the samples down to 5 K. In Figs. \ref{fig_Ag}(a) and \ref{fig_Bg}(a), we display the temperature dependence of the A$_{g}$ peaks and B$_{g}$ peaks, respectively. As expected, most peaks are slightly blue shifted and become a little sharper with temperature decreasing, except for the A$_g$ peak at 89.9~cm$^{-1}$ that we discuss below. We show in Fig. \ref{fig_Ag}(b), a quantitative analysis of the peak position and the linewidth of the A$_g$ Raman peak at 128.8~cm$^{-1}$, which has been fitted with a Lorentzian function. The results of similar analysis are also shown in Figs. \ref{fig_Bg}(b) and \ref{fig_Bg}(c) for the B$_g$ mode at 61.6~cm$^{-1}$. The peak position $\omega_{ph}(T)$ and the linewidth $\Gamma_{ph}(T)$ are consistent with simple expressions corresponding to the anharmonic phonon decay into acoustic phonons with the same frequencies and opposite momenta \cite{formular1,formular2}:

\begin{equation}
\label{eq_omega}
\omega_{ph}(T)=\omega_{0}-C\left( 1+\frac{2}{e^{\frac{\hbar\omega_0 }{ 2k_{B}T}} -1} \right )
\end{equation}
\begin{equation}
\label{eq_gamma}
\Gamma_{ph}(T)=\Gamma_{0}+\Gamma\left(1+\frac{2}{e^{\frac{\hbar\omega_0 }{2k_{B}T}} -1} \right ),
\end{equation}

\noindent where $C$ and $\Gamma$ are positive constants, $\omega_0$ is the bare phonon frequency, and $\Gamma_0$ is a residual, temperature-independent linewidth. From the fits, we extract $\omega_0=137.1$~cm$^{-1}$, $C=0.1755$~cm$^{-1}$, $\Gamma_0=3.496$~cm$^{-1}$ and $\Gamma=0.05918$~cm$^{-1}$ for the A$_{g}$ peak at 128.8~cm$^{-1}$. Similarly, the position and linewidth of the B$_{g}$ peak at 61.6~cm$^{-1}$ is well fitted with the formulas with $\omega_0=64.96$~cm$^{-1}$, $C=0.153$~cm$^{-1}$, $\Gamma_0=1.614$~cm$^{-1}$ and $\Gamma=0.02442$~cm$^{-1}$.

Ignoring the A$_g$ peak at 89.9~cm$^{-1}$, we could conclude in the absence of CDW this system. However, as shown in Fig. \ref{fig_Ag}(a), the linewidth of the A$_g$ peak at 89.9~cm$^{-1}$ fit with a single Lorentzian function exhibits an unusual increase, almost exponential, upon cooling. To emphasize this point, we show in the inset of Fig. \ref{fig_Ag}(c) the spectra of this peak at 5 K obtained from two different samples and two different laser excitations. Interestingly, the peak splits into two peaks at low temperature, which can explain the unusual increase of the linewidth observed upon cooling. It is important to notice that the peak at 89.9~cm$^{-1}$ has the A$_g$ symmetry and is by definition non-degenerate. In addition, no other peak that we did not observe is expected around that energy. The observation of a new peak is thus most likely caused by a lowering of the symmetry of the system. In that sense, the new peak emerging at low temperature may be related to the CDW-like gap feature observed at 4 K by STM \cite{STS}. Indeed, the A$_g$ phonon at 89.9~cm$^{-1}$ represents the vibration of Te$^1$, Te$^2$, Te$^5$, Te$^6$, Te$^7$ and Te$^8$, among which are atoms located at the bright and dark stripes observed by Du \emph{et al.} in STM measurements \cite{STM}, where the CDW-like gap feature is detected. If this is really the origin of the peak splitting, we estimate that the CDW transition temperature or the temperature at which CDW fluctuations emerge is somewhere between 140 K and 200 K, temperature range where the linewidth of the 89.9~cm$^{-1}$ starts to increase. Unfortunately, the proximity of the two-peak components prevents us from performing a reliable fit of the 89.9~cm$^{-1}$ feature using two Lorentizian functions at such high temperature.


In summary, we have performed polarized Raman scattering measurements on the newly discovered superconductor Ta$_{4}$Pd$_{3}$Te$_{16}$($T_c = 4.6$ K). We observe twenty-eight out of thirty-three Raman active  modes, with frequencies in good accordance with our first-principles calculations. Most of the phonons that we observe vary only slightly with temperature except for one Ag phonon at 89.9~cm$^{-1}$. The linewidth of the A$_{g}$ phonon mode at 89.9~cm$^{-1}$ shows an unconventional increase with temperature decreasing, which is possibly due to a CDW transition or to CDW fluctuations.

We acknowledge J. Ma, P. Zhang and J.-X. Yin for useful discussions. This work was supported by grants from MOST (2010CB923000, 2011CBA001000, 2011CBA00102, 2012CB821403 and 2013CB921703) and NSFC (11004232, 11034011/A0402, 11234014, 11274362 and 11474330) from China.

\bibliography{citation}

\end{document}